# On the group-theoretical approach to relativistic wave equations for arbitrary spin

Luca Nanni

*luca.nanni@edu.unife.it*

Formulating a relativistic equation for particles with arbitrary spin remains an open challenge in theoretical physics. In this study, the main algebraic approaches used to generalize the Dirac and Kemmer–Duffin equations for particles of arbitrary spin are investigated. It is proved that an irreducible relativistic equation formulated using spin matrices satisfying the commutation relations of the anti-de Sitter group leads to inconsistent results, mainly as a consequence of violation of unitarity and the appearance of a mass spectrum that does not reflect the physical reality of elementary particles. However, the introduction of subsidiary conditions resolves the problem of unitarity and restores the physical meaning of the mass spectrum. The equations obtained by these approaches are solved and the physical nature of the solutions is discussed.

## 1. Introduction

Formulating the Klein–Gordon equation [1] for spin-zero particles and the Dirac equation [2] for spin-half particles, together with discovering the positron [3] and neutron [4], gave strong impetuous to formulating a generalized relativistic equation for particles with arbitrary spin. Physicists of that period were convinced that shortly afterwards other particles with different masses and spins would be discovered, although theoretical physics then had no theory that could predict the existence of such particles. The first to confront this issue was Majorana in 1932 [5], although his main goal was to prove that the negative-energy solutions of the Dirac equation were not physically acceptable. Even though Majorana's goal turned out to be futile (a few months after the publication of his work, the positron was discovered), his research led to the first infinite-dimensional representation of the homogeneous Lorentz group. In that work, which was the precursor to his symmetric theory of the electron and positron published in 1937 [6], Majorana used group theory for the first time, which in the 1930s was still considered primarily part of pure mathematics. The Majorana equation provides a wide spectrum of solutions, including space-like ones. Moreover, for time-like solutions, an unexpected discrete mass spectrum arises. We reason that, apart from its mathematical interest, the Majorana equation may be useful for investigating the nature of particle masses [7] and for studying composite quantum systems with many degrees of freedom [8–10].

Majorana's work was rediscovered in the second half of the 1930s by Wigner [11] and Pauli [12]. It is also believed that Dirac read Majorana's paper before publishing the results of his study on relativistic wave equations for arbitrary spin, although he does not mention Majorana in his paper [13]. In fact, in his work, Dirac constructs the spin matrices following a Lie-algebra-based approach that recalls the approach used by Majorana in 1932, even though Dirac presents it using a more sophisticated formalism (that of spinors and tensors). Bhabha, to whom Pauli reported Majorana's work in the early 1940s, published in 1945 an article on the formulation of a relativistic equation for arbitrary spins by faithfully following Majorana's theory, except for the requirement of indeterminacy of the sign of energy [14]. In doing so, Bhabha recovered the solutions with negative energy (antiparticle states), eliminating the mathematical complexity arising from the need to solve an infinite system of differential equations. Bhabha's theory remains one of the most promising approaches to describing particles with spin greater than one. Not surprisingly, the equations of Kemmer and Duffin [15,16] for spin-one particles, of Rarita and Schwinger [17] for spin-three-halves particles and of Bargmann and Wigner [18] for particles with arbitrary spin can be obtained from that of Bhabha by introducing suitable subsidiary conditions [19].

During the last decades, the Standard Model of particle physics has evolved to one of the most precise theories in physics, describing the properties and interactions of fundamental particles in various experiments with high accuracy.

However, it lacks some shortcomings from the experimental as well as from the theoretical point of view. There is no approved mechanism for the generation of masses of the fundamental particles, in particular for massive neutrinos. Besides, the standard model does not explain the observance of dark matter in the universe. Moreover, the gauge couplings of the three forces in the standard model do not unify, implying that a fundamental theory combining all forces cannot be formulated. In this scenario, we address the relativistic theory of particles with arbitrary spin as an answer to these questions, emphasizing the problem of the mass spectrum of baryons and mesons.

In the following, we investigate and compare (in reverse chronological order) the different physical and mathematical approaches used by Bhabha, Dirac and Majorana for formulating relativistic wave equations. This allows us to understand better why a generalized relativistic equation cannot be obtained strictly within the framework of the Lie group of symmetries of spacetime, but instead requires the introduction of subsidiary conditions that lead to results for which an appropriate physical interpretation is needed in terms of known elementary particles. For each approach, we solve the obtained equations, using complex analysis when algebraic methods lead to excessive computational difficulties. We find that the generalized equations are nothing but finite or infinite systems of linear equations, each related to a given representation of the Lorentz group. The properties of the solutions obtained are also investigated. Finally, we propose some physical interpretations of the mass spectra, considering both their correlation with Sidharth's empirical formula and the possibility that they are related to composite systems formed by low-spin elementary particles interacting at a scale not far above their masses.

**2. The Bhabha approach**

Bhabha formulated a linear equation of the type

$$(i\hbar \Gamma^0 \partial_t - i\hbar c \Gamma^\mu \partial_\mu - \chi)\psi = 0, \tag{1}$$

where $\Gamma^\nu$ ($\nu = 0, 1, 2, 3$) are finite square matrices that satisfy the commutation relations of the Lorentz group in five dimensions. This group is recognized as the anti-de Sitter group O(3,2) and is obtained adding a timelike direction to the four-dimensional Minkowski spacetime [20]. O(3,2) has two timelike directions and three spacelike directions and, therefore, is not a spacetime in the ordinary sense (a Lorentzian manifold with one temporal and three spatial dimensions). However, the hypersurface with equation $(x_0{}^2 - x_1{}^2 - x_2{}^2 - x_3{}^2 + x_4{}^2) = R^2$ is a spacetime. It has constant positive curvature and reproduces (after a renormalization) the Minkowski spacetime when the curvature tends to zero. In this group, boosts and space rotations are transformations that leave the fifth axis unchanged. Therefore, the irreducible representations of the $\Gamma^\nu$ matrices are obtained from the irreducible representations of the O(3,2) group. The wave function $\psi(\boldsymbol{x}, \sigma)$ is a function of the spacetime coordinate $\boldsymbol{x} = (ct, x, y, z)$ and the spin coordinate $\sigma$. The term $\chi$ is related to an unspecified rest-mass energy and its physical meaning is clarified later in this work. The explicit form of this energy is given by the usual formula $\chi = mc^2$, but as discussed in §5, the mass $m$ depends on both the particle spin and a constant. Bhabha required no other subsidiary conditions for his theory.

Equation (1) must transform in a covariant manner under any transformations of the homogeneous Lorentz group. Denoting by $\Lambda_r^\nu$ a Lorentz transformation, we can rewrite (1) as

$$[i\hbar\Gamma^0(\Lambda^{-1}\Lambda)\partial_t - i\hbar c\Gamma^\mu(\Lambda^{-1}\Lambda)\partial_\mu - \chi]\psi = 0, \qquad (2)$$

which, on rearrangement, gives

$$[i\hbar(\Gamma^0\Lambda^{-1})(\Lambda\partial_t) - i\hbar c(\Gamma^\mu\Lambda^{-1})(\Lambda\partial_\mu) - \chi]\psi = 0. \qquad (3)$$

If we set

$$\Gamma^\nu\Lambda^{-1} = \Lambda\Gamma^\nu = (\Gamma^\nu)' \quad \text{and} \quad \Lambda\partial_\nu = \partial_\nu\Lambda^{-1} = (\partial_\nu)', \qquad (4)$$

then (1) becomes

$$[i\hbar(\Gamma^0)'(\partial_t)' - i\hbar c(\Gamma^\mu)'(\partial_\nu)' - \chi]\psi = 0. \qquad (5)$$

Therefore, the Lorentz transformation $\Lambda_r^\nu$ transforms the differential four-vector operator $\partial_\nu$ in $(\partial_\nu)'$ and the spin four-vector operator $\Gamma^\nu$ in $(\Gamma^\nu)'$. This means that there exists an operator $U$ such that

$$(\Gamma^r)' = U\Gamma^r U^{-1} \Rightarrow \Gamma^\nu = \Lambda_r^\nu U\Gamma^r U^{-1}. \tag{6}$$

The operators $U$ form a representation of the Lorentz group (which is a Lie group, i.e. a group of transformations that depend continuously on some parameters) of dimension $d$, whose kernel is given, for infinitesimal transformations, by the six generators $\boldsymbol{J} = (J_1, J_2, J_3)$ and $\boldsymbol{K} = (k_1, k_2, k_3)$ that satisfy the following commutation relations [21]:

$$[J_i, J_j] = i\varepsilon_{ijk}J_k, \quad [J_i, K_j] = i\varepsilon_{ijk}K_k, \quad [K_i, K_j] = -i\varepsilon_{ijk}J_k, \tag{7}$$

where $\varepsilon_{ijk}$ is the Levi-Civita three-index symbol. In relations (7) we recognize the quantum algebra of angular momentum. Therefore, the generators $J_i$ and $K_j$ are infinitesimal rotations of Minkowski spacetime (i.e. they differ little from the identity), thereby allowing the elements of the Lie group to be written in parametric form as $g(\alpha) = e^{i\alpha_a T^a}$, where $\alpha_a$ are real numbers and $T^a$ are the aforementioned generators. The dimension $d$ of the representation and therefore that of the matrices $\Gamma^\nu$ depend on the particle spin. By this algorithm, using the spin matrices for $s_0 = 1/2$ and/or $s_0 = 1$, we construct all those for spin $s = s_0 + 1$. Let us look at this method more closely. It is known that a Lorentz representation in anti-de Sitter space is characterized by two positive numbers [22], the first of which is the spin $s$ and the second, denoted by $m$, is a number given by $m = s, s-1, s-2, \ldots, 0$, where $m = 0$ occurs only for integer spin. Therefore, the number $m$ represents the projection of the spin along a given direction, such as the $z$ axis. Each pair $(s, m)$ identifies an irreducible representation of the O(3,2) group whose dimension is given by [14]

$$d_5(s, m) = \frac{2}{3}\left(s + \frac{3}{2}\right)\left(m + \frac{1}{2}\right)(s - m + 1)(s + m + 2). \tag{8}$$

If we project the five-dimensional anti-de Sitter space along the fifth axis, then we obtain a four-dimensional real space whose orthogonal group $O(3,1)$ has an irreducible representations of dimension [14]

$$\begin{cases} d_4(s,m) = 2(s - m + 1)(s + m + 1) & \text{if } m \neq 0, \\ d_4(s,m) = (s + 1)(s + 1) & \text{if } m = 0. \end{cases} \quad (9)$$

Equations (8) and (9) are tensorial representations of the $O(3,2)$ group obtained by an appropriate two-row Young diagrams. This (pictorial) formalism is a one-to-one correspondence with irreducible representations of the symmetric group over the complex numbers [23]. Therefore, having set the spin $s$, we obtain $s + 1/2$ equations of type (1) if $s$ is half of an odd integer (half-odd integer) or $s + 1$ equations if $s$ is an integer. These equations, which are nothing but linear systems of differential equations, are formed by $d_4(s,m)$ numbers of single coupled differential equations. For instance, if $s = 1$, then we have two equations of the type (1), the first formed by $d_4(1,1) = 6$ first-order differential equations and the second by $d_4(1,0) = 4$ first-order differential equations. Given that Bhabha modified the Majorana theory by reintroducing the negative-energy solutions, it is clear that the matrices $\Gamma^\nu$ are $d_4(s,m) \times d_4(s,m)$ square matrices, which in turn are formed from $\frac{1}{2}d_4(s,m) \times \frac{1}{2}d_4(s,m)$ spin matrices. The latter are obtained from angular momentum ladder operators constructed using suitable linear combinations of the generators $\boldsymbol{J}$ and $\boldsymbol{K}$ [23, 24]. Also, the matrices representing the six generators have dimensions that depend on the choice of spin. In particular, they are diagonal block matrices of the type $J_{ij}(s) = \text{diag}[J_{ij}(s,s), J_{ij}(s, s-1), \ldots]$ and $K_{ij}(s) = \text{diag}[K_{ij}(s,s), K_{ij}(s, s-1), \ldots]$. The components of the spin matrices forming the $\Gamma^\nu$ operators are

$$\begin{cases} (S_x)_{ij} = \dfrac{\hbar}{2}(\delta_{i,j+1} + \delta_{i+1,j})\sqrt{(s+1)(i+j+1) - ij}, \\ (S_y)_{ij} = i\dfrac{\hbar}{2}(\delta_{i,j+1} - \delta_{i+1,j})\sqrt{(s+1)(i+j-1) - ij}, \\ (S_z)_{ij} = \hbar\delta_{i,j}(s+1-i) \quad \text{if } i = j, \quad (S_z)_{ij} = 0 \quad \text{if } i \neq j, \end{cases} \quad (10)$$

where the indices $i$ and $j$ run from zero to $2s + 1$. The matrices $S_x$ and $S_y$ have non-trivial elements on the two diagonals parallel to the main one, while the matrix $S_z$ has non-trivial elements only on the main diagonal.

The obtained results clarify why the term $\chi$ in (1) is not defined unambiguously, at least not for spin-half and spin-one particles. In fact, the commutation relations of the anti-de Sitter group [25] for $s > 1$ lead to a $\Gamma^0$ matrix with more than two distinct eigenvalues. These commutation relations can be written as

$$[\Gamma^m, \Gamma^n] = \kappa I^{mn}, \tag{11}$$

where $I^{mn}$ is a generator of the anti-de Sitter group and $\kappa$ is a numerical constant. As suggested by Bhabha, this constant can be removed by multiplying by $\kappa^{-1/2}$, but this change is reflected in the value of the rest-mass energy $\chi$. In fact, rewriting (1) in the centre-of-mass reference frame,

$$(i\hbar \Gamma^0 \partial_t - \chi)\psi = 0, \tag{12}$$

and accounting for the fact that the matrix $\Gamma^0$ has distinct eigenvalues, which depend on the spin, we obtain a discrete mass spectrum. In particular, if $s$ is an integer, then we obtain $2s$ distinct eigenvalues of which half are positive and the other half are negative, whereas if $s$ is a half-odd integer, then we obtain $2s + 1$ distinct eigenvalues of which half are positive, and the other half are negative. Therefore, having set the particle spin, the Bhabha equation returns a discrete mass spectrum formed by fractional values of $\chi$, of which those with a positive (resp. negative) sign are attributed to particle (resp. antiparticle) states.

We can summarize with an example what has been argued so far. For a particle with spin $s = 3/2$ we have two equations (systems) of the type (1), namely two distinct irreducible representations of dimensions $d_4\left(\frac{3}{2},\frac{3}{2}\right) = 8$ and $d_4\left(\frac{3}{2},\frac{1}{2}\right) = 12$, respectively. This means that the first system comprises eight single differential equations and the second twelve. For the mass spectrum, we have four values: $m = \pm\frac{2}{3}\chi$ and $m = \pm\frac{2}{3}\chi$. The first pair of masses is attributed to the irreducible representation of dimension $d_4\left(\frac{3}{2},\frac{3}{2}\right)$ and the second to the representation of dimension $d_4\left(\frac{3}{2},\frac{1}{2}\right)$.

Whatever the chosen spin, the Bhabha equation is equivalent to a finite system of differential equations whose coefficient matrix is non-singular. In fact, all spin matrices $\Gamma^\nu$ are invertible. Therefore, the system admits only one solution, which can be calculated using Cramer's rule. For free particles, the solutions are wave functions of the type

$$\psi = \varphi(E, \mathbf{p})\, e^{\mp i(p_x x + p_y y + p_z z - Et)/\hbar}, \tag{13}$$

where $\varphi(E, \mathbf{p})$ is a spinor whose non-trivial components are suitable combinations of the four-momentum components $(E/c, p_x, p_y, p_z)$, while $E$ is the particle energy, whose value depends on the spin, namely

$$E^2 = p^2 c^2 + (\chi/s)^2. \tag{14}$$

where $p^2$ is the dot product of the Euclidean vector $\mathbf{p} = (p_x, p_y, p_z)$. Therefore, the energy $E$ depends on the spin via the term $\chi$. Substituting (13) into (1) and performing all the derivatives, we obtain a system of algebraic equations in which the unknowns are the spinor components $\varphi_\nu$:

$$[i\hbar(\Gamma^{\rho\nu})^0 \varphi_\nu \partial_t - i\hbar c(\Gamma^{\rho\nu})^\mu \varphi_\nu \partial_\mu - \chi\varphi_\nu] e^{\mp i(p_x x + p_y y + p_z z - Et)/\hbar} = 0. \tag{15}$$

Using Cramer's rule, we obtain all the values of $\varphi_\nu$, namely the complex numbers $0, 1, \alpha[cp_z/(E + \chi/s)]^n$, and $\beta[c(p_x \pm ip_y)/(E + \chi/s)]^n$, where $n = s$ for integer spin and $n = s + 1/2$ for half-odd integer spin, while $\alpha$ and $\beta$ are constant numbers from among the elements of the spin matrices. For clarity, let us reconsider the example of a spin-three-halves particle. For the particle state whose spin is up and has a $z$-component of $3/2$, the spinor $\varphi(E, \mathbf{p})$ can be written as

$$\varphi(E, \mathbf{p}) = (1, 0, 0, 0, \varphi_5, \varphi_6, \varphi_7, \varphi_8)^T e^{\mp i(p_x x + p_y y + p_z z - Et)/\hbar}. \tag{16}$$

Substituting this spinor into the Bhabha equation for $s = \frac{3}{2}$ corresponding to the irreducible representation of dimension $d_4\left(\frac{3}{2},\frac{3}{2}\right) = 8$, we obtain

$$\begin{cases} \varphi_5, \varphi_6 \propto \left[\dfrac{cp_z}{E + 2\chi/3}\right]^2, \\ \varphi_7, \varphi_8 \propto \left[\dfrac{c\left(p_x \pm ip_y\right)}{E + 2\chi/3}\right]^2. \end{cases} \qquad (17)$$

This procedure must be repeated for all other possible particle–antiparticle states with spin up and spin down.

The main inconsistency of Bhabha's theory arises from the fact that the representations of the homogeneous Lorentz group are finite and therefore non-unitary, at least for all spin values greater than one. Furthermore, each component of the wave function by itself does not satisfy the Klein–Gordon equation, which it should do in any relativistic quantum theory. In addition to these inconsistencies, we are also dealing with a mass spectrum that cannot be interpreted on the basis of current knowledge of particle physics. We postpone our discussion of these aspects to §6.

## 3. The Dirac approach

Dirac constructed a relativistic wave equation for spin greater than $\frac{1}{2}$ such that the representations of the Lorentz group were unitary and each component of the wave function by itself satisfied the Klein–Gordon equation [13]. The main problem with this approach is that unitary representations require spin matrices that do not ensure the Lorentz invariance of the equation. Dirac overcame this difficulty by the *ad hoc* introduction of two auxiliary matrices, although these complicate the mass spectrum that is obtained, which depends on two parameters rather than just one.

To formulate his theory, Dirac wrote down the generic spin operator as a six-vector of components $(S_{xt}, S_{xy}, S_{xz}, S_{yt}, S_{yz}, S_{zt})$, in units of $\hbar$, These components satisfy the angular momentum commutation relations [26]. From these components, it is possible to construct six generators $\alpha_\mu$ and $\beta_\mu$, with $\mu = 1,2,3$, that behave like independent angular momenta in $\mathbb{R}^3$:

$$\alpha_\mu = \frac{1}{2}(S_{jk} - iS_{it}), \qquad \beta_\mu = \frac{1}{2}(S_{jk} + iS_{it}), \qquad (18)$$

where the six operators are obtained by cyclic permutations of $i, j, k$ from $x$ to $z$. The operators (18) satisfy the commutation relations

$$\begin{cases} \alpha_i \alpha_j - \alpha_j \alpha_i = i\alpha_k, \\ \beta_i \beta_j - \beta_j \beta_i = i\beta_k, \end{cases} \qquad (19)$$

in addition to the following relations involving the squares of the individual components:

$$\begin{cases} \sum_{\mu=1}^{3} \alpha_\mu^2 = k(k+1), \\ \sum_{\mu=1}^{3} \beta_\mu^2 = l(l+1). \end{cases} \qquad (20)$$

Here, $k$ and $l$ are integers or half-odd integers and are the two parameters that characterize the representations of the homogeneous Lorentz group. As expected, they coincide with those obtained from Bhabha's theory. Equations (20) show that $\alpha_\mu$ and $\beta_\mu$ are square matrices of dimensions $k(k+1) \times k(k+1)$ and $l(l+1) \times l(l+1)$, respectively. From these matrices, we construct two new operators

$$A = \begin{pmatrix} \alpha_x & \alpha_x - i\alpha_y \\ \alpha_x + i\alpha_y & -\alpha_x \end{pmatrix}, \qquad B = \begin{pmatrix} \beta_x & \beta_x - i\beta_y \\ \beta_x + i\beta_y & -\beta_x \end{pmatrix}, \qquad (21)$$

which satisfy

$$A(A+1) = k(k+1), \qquad B(B+1) = l(l+1) \qquad (22)$$

and whose eigenvalues are $k, k-1, \ldots, -k$ and $l, l-1, \ldots, -l$, respectively. The dimension of the matrices $A$ and $B$ are $2(2k+1) \times 2(2k+1)$ and $2(2l+1) \times 2(2l+1)$, respectively, i.e. they have twice as many elements as the matrices $\alpha_\mu$ and $\beta_\mu$. In this, Dirac faithfully followed the approach used to formulate his equation for spin-$\frac{1}{2}$ particles, where the matrices $\alpha_\mu$ have twice the dimension of the Pauli matrices from which they are formed [2].

The next step to find a unitary transformation $U$ that makes the matrices $A$ and $B$ diagonal. We should note that no requirement has been imposed regarding the algebraic nature of the matrices $\alpha_\mu$ and $\beta_\mu$. Specifically, it is not required that these matrices be Hermitian (which would ensure the unitarity of the representations of the Lorentz group). The transformation that diagonalizes the matrices $A$ and $B$ is

$$\begin{cases} U^{-1}AU = \text{diag}(q \text{ values of } k, p \text{ values of } -(k+1)), \\ U^{-1}BU = \text{diag}(q \text{ values of } l, p \text{ values of } -(l+1)), \end{cases} \quad (23)$$

where $q$ and $p$ are the multiplicities of the eigenvalues of $A$ and $B$. Dirac proved that his theory was consistent only if

$$\begin{cases} m = 2(k+1), 2(l+1), \\ n = 2k, 2l. \end{cases} \quad (24)$$

where $m$ is the number of columns of matrix $A$ and $n$ is the number of columns of matrix $B$. The relations (24) satisfy the constraint $m + n = 2(2k+1)$ and ensure that the matrices $\alpha_z$ and $\beta_z$ have eigenvalues $k, \ldots, -k$ and $l, \ldots, -l$, respectively. We have thus obtained explicit forms for the matrices $\alpha_z$ and $\beta_z$ through which, using the (19) and (20), we obtain those for the matrices $\alpha_x, \alpha_y$ and $\beta_x, \beta_y$.

We now have all the algebraic tools for writing down the relativistic equation for a particle of spin $s$. Since the matrices $A$ and $B$ have eigenvalues with different multiplicities, Dirac split the wave function $\psi$ into two parts, $\psi_A$ and $\psi_B$, of which the former has $(2k+1); 2l$ components and the latter $(2l+1); 2k$ components. The wave functions $\psi_A$ and $\psi_B$ must satisfy the coupled equations

$$\begin{cases} (i\hbar\mathbb{1}\partial_t - i\hbar c\alpha^\mu\partial_\mu)\psi_A = m'c^2\delta\psi_B, \\ (i\hbar\mathbb{1}\partial_t - i\hbar c\beta^\mu\partial_\mu)\psi_B = m''c^2\gamma\psi_A, \end{cases} \quad (25)$$

where $\delta$ and $\gamma$ are non-singular matrices of dimensions $2k \times 2k$ and $2l \times 2l$, respectively. These matrices have been introduced in an *ad hoc* manner so that the Klein–Gordon equation is satisfied. In fact, obtaining $\psi_B$ from the first equation in (25) and replacing it in the second one, we get

$$\left(-\hbar^2 \mathbb{1} \partial_t^2 + \hbar^2 \beta^\mu \alpha_\mu \partial_\mu^2\right)\psi_A = m'm''c^4 \gamma\delta\psi_A \qquad (26)$$

which is the Klein–Gordon equation if we impose $\beta^\mu \alpha_\mu = \mathbb{1}$ and $\gamma\delta = \mathbb{1}$. These algebraic constraints represent the subsidiary conditions needed to make the relativistic equation consistent with the typical energy-momentum relation. The term $m'm''c^4$ can only have the meaning of the squared rest mass energy and, as the matrices $A$ and $B$ have been constructed, depends on the two parameters $k$ and $l$. In fact, setting $m'm''c^4 = \chi^2$ (to use the same formalism as in Bhabha's theory), we easily get the following relations:

$$m'c^2 = \chi(k/l)^{1/2} \quad \text{and} \quad m''c^2 = \chi(l/k)^{1/2} \qquad (27)$$

Therefore, the two mass spectra are duals. From (27), it can be seen that the matrices $\delta$ and $\gamma$ can be chosen as $\text{diag}[(k/l)^{1/2}]$ and $\text{diag}[(l/k)^{1/2}]$. In this way, their product trivially gives the unit matrix, as (26) requires.

Compared with Bhabha's theory, the mass spectrum that arises from (25) is characterized by two parameters. For instance, a particle with rest energy $\chi$ and spin $s = \frac{3}{2}$ presents the following mass states:

$$\begin{cases} (3/2,3/2) \Rightarrow \psi_A \to m' = \dfrac{\chi}{c^2} \text{ and } \psi_A \to m'' = \dfrac{\chi}{c^2}, \\ (3/2,1/2) \Rightarrow \psi_A \to m' = \dfrac{\sqrt{3}\chi}{c^2} \text{ and } \psi_A \to m'' = \dfrac{\chi}{\sqrt{3}c^2}, \\ (3/2,-1/2) \Rightarrow \psi_A \to m' = -\dfrac{\sqrt{3}\chi}{c^2} \text{ and } \psi_A \to m'' = -\dfrac{\chi}{\sqrt{3}c^2}, \\ (3/2,3/2) \Rightarrow \psi_A \to m' = -\dfrac{\chi}{c^2} \text{ and } \psi_A \to m'' = -\dfrac{\chi}{c^2}, \end{cases} \qquad (28)$$

where the negative values refer to antiparticle states. As expected, with the introduction of the matrices $\delta$ and $\gamma$ into the mass terms of (25), the obtained mass spectrum is twice that of Bhabha's theory. The number of differential equations that form the coupled system (25), on the other hand, remains unchanged compared with the case $s = \frac{3}{2}$ addressed in the previous section.

The system of coupled equations (25) can be solved easily using Cramer's rule, considering that the coefficient matrix is invertible. However, the spinors $\psi_A$ and $\psi_B$ will have to be further split into two parts, since each of them contains terms referring to two distinct mass values. Remaining with the example $=\frac{3}{2}$, we can write

$$\begin{cases} \psi_A = \psi_A{'}(\pm\chi/c^2)\oplus\psi_A{''}(\pm\sqrt{3}\chi/c^2), \\ \psi_B = \psi_B{'}(\pm\chi/c^2)\oplus\psi_B{''}(\pm\sqrt{3}\chi/c^2), \end{cases} \quad (29)$$

where the explicit forms of $\psi_A{'}, \psi_A{''}, \psi_B{'}$ and $\psi_B{''}$ are

$$\begin{cases} \psi_A{'} = (\varphi_1, \varphi_2)^{\mathrm{T}} e^{\mp i(p_x x + p_y y + p_z z - E't)/\hbar}, \\ \psi_A{''} = (\varphi_3, \varphi_4)^{\mathrm{T}} e^{\mp i(p_x x + p_y y + p_z z - E''t)/\hbar}, \\ \psi_B{'} = (\varphi_5, \varphi_6)^{\mathrm{T}} e^{\mp i(p_x x + p_y y + p_z z - E'''t)/\hbar}, \\ \psi_B{''} = (\varphi_7, \varphi_8)^{\mathrm{T}} e^{\mp i(p_x x + p_y y + p_z z - E''''t)/\hbar}. \end{cases} \quad (30)$$

The energies $E', \dots, E''''$ are given by

$$\begin{cases} E' = E''' = (p^2 c^2 + \chi^2)^{1/2}, \\ E'' = \left[p^2 c^2 + (\sqrt{3}\chi)^2\right]^{1/2}, \quad E'''' = \left[p^2 c^2 + (\chi/\sqrt{3})^2\right]^{1/2}. \end{cases} \quad (31)$$

Successive substitution of (31) into (30), (30) into (29) and (29) into (25) gives two systems, each of four algebraic equations, from which we obtain the eight spinor components. Overall, the method of solving (25) is similar to that discussed for the Bhabha equation, only a little more laborious. Depending on the state under consideration, the non-trivial components of the four-spinor will have a form similar to (17). For instance, for the spin-up state $\left(\frac{3}{2}, \frac{1}{2}\right)$, we have

$$\begin{cases} \varphi_5 \propto \left[\dfrac{cp_z}{E+\chi}\right]^2, \quad \varphi_6 \propto \left[\dfrac{c(p_x \pm ip_y)}{E+\sqrt{3}\chi}\right]^2, \\ \varphi_7 \propto \left[\dfrac{cp_z}{E+\chi}\right]^2, \quad \varphi_8 \propto \left[\dfrac{c(p_x \pm ip_y)}{E+\chi/\sqrt{3}}\right]^2. \end{cases} \quad (32)$$

## 4. The Majorana approach

Majorana's approach was aimed at eliminating the negative-energy solutions that appear in the Dirac equation for spin-$\frac{1}{2}$ particles [5]. This was in 1932, and when Majorana formulated his theory, the positron had yet to be discovered (or at least the news of its discovery had not yet arrived in Italy). Therefore, Majorana considered these solutions to be unphysical and that the eigenvalues of the energy operator should not be square roots of real numbers. Majorana's aim, therefore, was to find a relativistic equation that admits only solutions with positive frequency and that, for slow motions, reduces to the Schrödinger equation. The only way to achieve this goal is to eliminate the uncertainty in the sign of the energy, by formulating an equation that admits only a single root. Algebraically, this is possible by considering infinite-component wave functions. The Majorana equation can be written as (1), but with $\Gamma^\nu$ obtained as infinite sequences of finite spin matrices. These matrices provide an infinite-dimensional representation of the homogeneous Lorentz group. Therefore, the root of the equation is a wave function with an infinite number of components that cannot be split into finite spinors, since is not possible to decide a priori to which spin a component corresponds.

The starting point of the Majorana approach is to require that the quadratic form $\psi^* \Gamma^0 \psi$ be positive-definite and that it transform covariantly under the action of the elements of the homogeneous Lorentz group. With these constraints, Majorana determined the structure of all the matrices $\Gamma^\nu$. In this section, we do not intend to investigate the algebraic method developed by Majorana: other authors have published detailed studies of the scientific importance of Majorana's work [24, 25, 28–31]. Rather, we want to explicitly determine and investigate all the possible solutions: time-like, light-like and space-like. In fact, unlike those of Bhabha and Dirac, the Majorana equation allows superluminal solutions with a continuous mass spectrum, a peculiarity attributed to the requirement of a positive-definite $\psi^* \Gamma^0 \psi$ term.

Let us write the Majorana equation as

$$\left(i\hbar\Gamma^0\partial_t - i\hbar c\Gamma^\mu\partial_\mu - \chi\right)\psi = 0, \tag{33}$$

where, as usual, $\chi$ is an undefined rest-mass energy. The wave function $\psi$ depends on the spacetime coordinates and on the infinite components of the spin space, which transform according to the unitary representation of the O(3,1) group generated by the Lie operators $\mathbf{J}$ and $\mathbf{K}$ [5]. The structure of the matrices $\Gamma^0$ and $\Gamma^\mu$ is very similar to those of Bhabha, with the non-trivial components of $\Gamma^\mu$ lying along the main diagonal and those of $\Gamma^\mu$ being blocks along one of the two diagonals. In particular, the non-trivial components of the matrix $\Gamma^0$ are given by $s + 1/2$. For convenience we rewrite (33) as

$$i\hbar\partial_t\psi = \mathbb{H}\psi, \tag{34}$$

where the Hamiltonian $\mathbb{H}$ is given by

$$\mathbb{H} = i\hbar c \sum_{\mu}^{3} \Gamma^\mu + \chi(\Gamma^0)^{-1}. \tag{35}$$

In (35) the matrix $(\Gamma^0)^{-1}$ has eigenvalues given by $1/(s + 1/2)$ with multiplicity $2s + 1$. Regarding the structure of the other three matrices, $\Gamma^3$ has non-trivial elements on the main diagonal, given by $(m, m - 1, \ldots, -m)$, while $\Gamma^2$ and $\Gamma^3$ have non-trivial elements on the two diagonals parallel to the main one. Taking the wave function as the plane wave (13), where the spinor $\varphi(E, \mathbf{p})$ has an infinite number of components and the energy $E$ is a function of the spin $s$, and considering the case of half-odd integer spin, we find that the Hamiltonian matrix takes a form in which the spin matrices with progressively increasing values develop along the main diagonal, giving the typical block structure shown in figure 1. By inserting this matrix into (34), we obtain a system of infinite algebraic equations with an infinite number of components. The matrix in figure 1 is clearly non-singular. This means that $\mathbb{H}$ has a non-zero determinant, a necessary but not sufficient condition for an infinite system to admit only one root. To ensure the uniqueness of the solution, the vector of coefficients, whose components are $i\hbar\partial_t\psi_i = E_i$, must be bounded [32]. The Majorana equation satisfies this condition, given that the rest-mass energy decreases with increasing spin

| $\chi - cp_z$ | $c(p_x + ip_y)$ | | | | | |
|---|---|---|---|---|---|---|
| $c(p_x - ip_y)$ | $\chi + cp_z$ | | 0 | | | |
| | | $(\chi - 3cp_z)/2$ | $\sqrt{3}c(p_x + ip_y)/2$ | 0 | 0 | |
| 0 | | $\sqrt{3}c(p_x - ip_y)/2$ | $(\chi - 3cp_z)/2$ | $(p_x + ip_y)/2$ | 0 | 0 |
| | | 0 | $(p_x - ip_y)/2$ | $(\chi + 3cp_z)/2$ | $\sqrt{3}c(p_x + ip_y)/2$ | |
| | | 0 | 0 | $\sqrt{3}c(p_x - ip_y)/2$ | $(\chi + 3cp_z)/2$ | |
| | | | 0 | | | |

**Figure 1**. Infinite Hamiltonian matrix for half-odd integer spin.

$$E^2(s) = p^2c^2 + \left(\frac{\chi}{s+\frac{1}{2}}\right)^2. \tag{36}$$

We have thus proved that (34) admits only one solution. However, the spinor components can be calculated through finite truncations [32]. Although laborious, this method is justified by the fact that the spinor components are given by

$$\begin{cases} \varphi_\mu^{(s)} \propto \left[\pm \dfrac{cp_z}{E+\chi}\right]^{s+1/2}, & \varphi_\nu^{(s)} \propto \left[\dfrac{c(p_x \pm ip_y)}{E+\chi}\right]^{s+1/2}, \\ \varphi_\mu^{(s)} \propto \left[\pm \dfrac{cp_z}{E+\chi}\right]^{s+1}, & \varphi_\nu^{(s)} \propto \left[\dfrac{c(p_x \pm ip_y)}{E+\chi}\right]^{s+1}, \end{cases} \tag{37}$$

where the first line holds for half-odd integer spin and the second line for integer spin. The indices $\mu$ and $\nu$ are related by $\nu = \mu + 1$. For time-like solutions, where the particle velocity is always lower than the speed of light, the quantities in square brackets in (37) are less than one and therefore, because these quantities are raised to the power of either $s + 1/2$ or $s + 1$, the spinor components become progressively smaller with increasing spin $s$. This explains why it is possible to solve the infinite system of equations by finite truncations. By choosing a cutoff beyond which the spinor components (37) are deemed insignificant, we can approximate the infinite system as a finite one that can be solved easily using Cramer's rule. However, this approach cannot be used for light-like and space-like solutions, given that the quantities in square brackets are then equal to or greater than one. This means that with increasing spin, the components in (37) retain the same order of magnitude (light-like solutions) or increase progressively (space-like solutions).

Given the difficulty in solving the Majorana equation with algebraic methods, we will use an analytical approach (despite other methods based on quaternionic algebra being effective options [32]) based on the fact that once a solution for a particular configuration has been obtained, it is possible to obtain all the others through Lorentz transformations. Suppose that $\psi(p_\nu)$ is a particular solution and we seek the solution for another configuration. Then, proceeding similarly to § 2, we find that there exists a matrix $U(\Lambda)$, which depends on the Lorentz transformation that relates the two configurations under consideration, such that

$$\psi'(p_\nu) = U(\Lambda)\psi(p_\nu) \quad \Rightarrow \quad \psi(p_\nu) = U^{-1}(\Lambda)\psi'(p_\nu). \tag{38}$$

Substituting this into (33) gives

$$\left(i\hbar\Gamma^0\partial_t - i\hbar c\Gamma^\mu\partial_\mu - \chi\right)U^{-1}(\Lambda)\psi'(p_\nu) = 0, \tag{39}$$

from which, by multiplying the left-hand side by $U(\Lambda)$, we obtain

$$\left(i\hbar U\Gamma^0 U^{-1}\partial_t - i\hbar c U\Gamma^\mu U^{-1}\partial_\mu - \chi\right)\psi'(p_\nu) = 0. \tag{40}$$

As $U\Gamma^\nu U^{-1}$ is equal to $\Lambda^\nu_r U\Gamma^r$, (40) shows that $\psi'(p_\nu)$ also satisfies (33). Let us first consider the case of time-like solutions. The simplest configuration from which to obtain a particular solution is that in the centre-of-mass reference frame, where $E = mc^2$ and $\boldsymbol{p} = 0$. The Majorana equation then becomes

$$(i\hbar\Gamma^0\partial_t - \chi)\psi = 0 \quad \Rightarrow \quad m = \frac{\chi}{c^2(s + 1/2)}, \tag{41}$$

since the eigenvalues of $\Gamma^0$ are $s + 1/2$. As anticipated above, the time-like solutions present a discrete mass spectrum. To understand the physical nature of these solutions, we introduce a representation of the anti-de Sitter group $O(3,2)$ through complex analysis [34–37]

$$z = x + iy, \quad \bar{z} = x - iy, \quad \partial = \frac{\partial}{\partial z}, \quad \bar{\partial} = \frac{\partial}{\partial \bar{z}}. \tag{42}$$

Using these transformations, we can write the four-vector $\Gamma^\nu$ as

$$\Gamma^\nu = \left[\frac{1}{2}(z\bar{z} - \partial\bar{\partial}), -\frac{1}{4}(z\bar{\partial} + z\partial), \frac{1}{4}i(z\bar{\partial} + \bar{z}\partial), -\frac{1}{2}(z\bar{z} + \partial\bar{\partial})\right]. \tag{43}$$

With this representation, we have moved to a two-dimensional space where the operators (42), normalized to make them dimensionless, take the following form

$$z = \frac{x + iy}{\lambda_0}, \qquad z\bar{z} = \frac{x^2 + y^2}{\lambda_0^2}, \qquad \partial\bar{\partial} = \frac{1}{4}\lambda_0\left(\frac{\partial^2}{\partial x^2} + \frac{\partial^2}{\partial y^2}\right), \qquad (44)$$

where $\lambda_0 = (x^2 + y^2)^{1/2}$ is a unit length. Substituting the time component of (43) into (41) and using (44), we obtain

$$\left[-\frac{\hbar^2}{2\mu}\left(\frac{\partial^2}{\partial x^2} + \frac{\partial^2}{\partial y^2}\right) + \frac{1}{2}\mu\omega^2\lambda^2 - E\right]\psi = 0, \qquad (45)$$

where

$$\omega = \frac{2\hbar}{\mu\lambda_0^2}, \qquad E = \frac{2\hbar^2}{\mu\lambda_0^2}. \qquad (46)$$

Equation (45) is the Schrödinger equation of a two-dimensional harmonic oscillator in an attractive parabolic potential. Therefore, the solution of the Majorana equation in the centre-of-mass reference frame is an infinite set of harmonic oscillators with reduced mass given by

$$\mu = \frac{1}{c^2}\frac{\frac{\chi}{s+\frac{1}{2}}\frac{\chi}{(s+\frac{1}{2})+1}}{\frac{\chi}{s+\frac{1}{2}} + \frac{\chi}{(s+\frac{1}{2})+1}} = \frac{\chi}{c^2 2(s+1)}. \qquad (47)$$

As the spin increases, the frequency and energy of the oscillator increase. Therefore, the solution of the Majorana equation is a function with an infinite number of components given by $\psi_s = h(x,y)\exp[-\chi\omega(x^2 + y^2)/4\hbar c^2(s+1)]$. The spin $s$ can be written as $2s(n_x + n_y)$, where $n_x$ and $n_y$ are the quantum numbers (integer numbers) of the two-dimensional oscillator. The function $h(x,y)$ depends on the quantum numbers $n_x$ and $n_y$ and is a polynomial function of the two spatial coordinates [37]. This approach shows that although the mass spectrum decreases with increasing spin, the mass states represented by the harmonic oscillators become progressively more energetic. This would seem to contradict experimental results, according to which lighter elementary particles are more stable than heavy ones (for example, the muon is heavier than the electron but rapidly decays into lighter and very stable particles). This contradiction would seem to compromise the application of the Majorana model for elementary particles, but at the same time suggests that it could be useful for the study of systems of interacting particles, not interpretable by the Standard Model, currently unknown or only hypothesized (such as dark matter or baryons exotic such as pentaquarks). If we had used the purely algebraic method of solution, this aspect of fundamental importance for the theory would not have emerged, and its absence would have led to a misleading interpretation of the mass spectrum.

Let us consider now the massive light-like solutions. The best configuration for the particular solution in this case is that in which the impulse has its non-zero component along the $z$ axis. The Majorana equation then becomes

$$(i\hbar \Gamma^0 \partial_t \pm i\hbar c \Gamma^3 \partial_z - \chi)\psi = 0. \tag{48}$$

Since the free particle is travelling at the speed of light, it is convenient to use the parametrization $(i\hbar \partial_t)\psi = (i\hbar c \partial_z)\psi = \hbar \omega$. Considering that the matrices $\Gamma^0$ and $\Gamma^3$ are diagonal, it is easy to obtain the energy spectrum associated with (48)

$$\hbar \omega = \frac{\chi}{\left(s + \frac{1}{2}\right) \pm \delta}, \tag{49}$$

where $\delta$ are the eigenvalues of $\Gamma^3$. The physical meaning of the particular solution becomes clear if we rewrite (48) using the complex operators (44)

$$\begin{cases} \left[-\frac{\hbar^2}{2\mu}\left(\frac{\partial^2}{\partial x^2} + \frac{\partial^2}{\partial y^2}\right) - \frac{2\hbar \chi}{\mu \omega \lambda_0^2}\right]\psi = 0, \\ \left(\hbar \omega \frac{x^2 + y^2}{\lambda_0^2} - \chi\right)\psi = 0, \end{cases} \tag{50}$$

where the first and second equations correspond to the $+$ and $-$ signs respectively in front of the $\Gamma^3$ term in (48) and $\omega = 2\hbar/\mu\lambda_0^2$. The first equation in (50) is the Schrödinger equation for a particle in a constant potential and with zero total energy. As usual, the spin dependence is contained in the reduced mass, as in the time-like case. Since a light-like particle cannot have zero energy, the obtained result can be interpreted by assuming that the first equation in (50) describes the motion of a composite system whose total energy is zero. The second equation, on the other hand, can be rewritten as

$$(x^2 + y^2)\psi = \frac{\lambda_0^2 \chi}{\hbar\omega}\psi \quad \Rightarrow \quad x^2 + y^2 = \frac{\lambda_0^2 \chi}{\hbar\omega}. \tag{51}$$

This is the Cartesian equation of a circle of radius $\lambda_0^2 \chi/\omega$, which implies that the motion of the harmonic oscillator associated with each spin is constrained to lie on a circle.

Finally, we consider space-like solutions, choosing the configuration to be that in which the particle has zero energy (a reference frame with infinite velocity) and only a $z$ component of spatial momentum. The Majorana equation then becomes

$$(i\hbar c \Gamma^3 \partial_z - \chi)\psi = 0 \quad \Rightarrow \quad m^2 = -(\chi/\delta c^2)^2, \tag{52}$$

where $\chi$ is an imaginary rest energy. In this case also, the physical meaning of the solution becomes clear when we rewrite (52) using the complex operators (44)

$$\left[-\frac{\hbar^2}{2\mu}\left(\frac{\partial^2}{\partial x^2} + \frac{\partial^2}{\partial y^2}\right) - \frac{1}{2}\mu\omega^2\lambda^2 - E\right]\psi = 0. \tag{53}$$

Equation (53) is completely analogous to (45) except for the presence of a repulsive parabolic potential. Therefore, the space-like solutions are energetically unstable but still possible. If instead the particle is subluminal with impulse $p_z$, then we obtain an equation completely similar to (53) but with an attractive parabolic potential. This proves that the time-like solutions, unlike the tachyon ones, are stable.

**5. Mass quantization**

The connection between the masses of elementary particles and the symmetries of the Lorentz group has long been a topic of debate [38–40]. According to Valmarov [41], the classification of relativistic wave equations is based on the interlocking representations of the Lorentz group. A system of interlocking representations is associated with a system of eigenvector subspaces of the energy operator. Such a correspondence allows defining matter spectrum, where each level of this spectrum presents some state of elementary particle. The theories investigated in the present study lead to a discrete mass spectrum, the correctness of which can be assessed only after a choice of the rest energy $\chi$ has been made. This is a feature shared by all quantum theories dealing with representations of the Lorentz group [40]. In our opinion, this is due to the difficulty in finding a unitary representation of the Lorentz group without introducing subsidiary conditions or other physical or mathematical constraints. Recently, an empirical formula has been proposed by Sidharth for the mass spectrum of baryons and mesons. It reads [41, 42]

$$m = 137 p \left(q + \frac{1}{2}\right), \tag{54}$$

where 137 is the pion mass in MeV/$c^2$, while $p$ and $q$ are positive integer numbers. The formula (54) reproduces the entire mass spectrum of known baryons and mesons with errors never exceeding 3%. Sidharth suggested that the numbers $p$ and $q$ could be quantum numbers of harmonic oscillators, without giving a satisfactory proof. Following this idea and considering that the time-like solutions of the Majorana equation can be interpreted in terms of the energy spectrum of harmonic oscillators with varying reduced mass, we can think of correlating the formula (54) with the expressions in (46). The wave equations for particles with arbitrary spin, in fact, are suitable for describing composite systems formed by elementary particles [8,25]. In the centre-of-mass frame, the Majorana oscillator energy can be rewritten as

$$E = m(s)c^2 = \frac{2\hbar^2}{\mu \lambda_0^2} \quad \Rightarrow \quad m(s) = \frac{2\hbar^2}{c^2 \mu \lambda_0^2}, \tag{55}$$

and, on substituting for $\mu$ from (47), we get the Majorana mass spectrum

$$m(s) = \frac{2\hbar^2 (s+1)}{\chi \lambda_0^2}. \tag{56}$$

By equating this expression and the formula (54), we get an algebraic relation between the spin $s$ and the positive integer numbers $p$ and $q$, besides providing physical meaning to the rest energy $\chi$

$$s = pq + \frac{1}{2}p - 1, \quad \chi = \frac{4\hbar^2}{137 \lambda_0^2}. \tag{57}$$

The value of the constant $\chi$ is $1.13687 \times 10^{-27}$ MeV/$\lambda_0^2$. If we assume that the order of magnitude of $\lambda_0$ is no greater than $10^{-15}$ m (i.e. the average dimension of a nucleus), then $\chi$ is of the order of 1 GeV. From the first expression in (57), we see that half-odd integer spin can be obtained only if $p$ is an odd positive integer. By combining all the possible values of $p$ and $q$, all spin sequences are obtained. Majorana theory, however, does not provide the numerical value of the constant $\chi$, which must be calculated using experimental data. This is also the case for the Standard Model, where, for example, the coupling constants are obtained from experimental data. In this framework, the mass of the pion takes the meaning of physical constant from which it is possible to get the entire mass spectrum of baryons and mesons.

It should be noted that the constant $\chi$ depends on the particular theory under consideration. In fact, we can also apply the complex operators (44) to the Bhabha and Dirac equations, obtaining finite sets of two-dimensional harmonic oscillators whose reduced mass depends on the algebra of the representation of the Lorentz group. For instance, in the Bhabha theory, where the energy spectrum is given by $\chi/s$, the reduced mass becomes $\mu = \chi/(2s+1)$, which leads to the formula $\chi = 2\hbar^2/137\lambda_0^2$ and spin $s = \frac{1}{2}pq + \frac{1}{4}p - \frac{1}{2}$. In the case of the Dirac theory, things get a little more complicated, since we have two dual mass spectra and therefore two values of $\chi$.

The interpretation we have given to the mass spectrum could be a little speculative and forced, but the logical connection between the empirical formula and the fact that the relativistic equations can always be reformulated as those of quantum harmonic oscillators is evident. Therefore, our speculation still has a robust scientific basis.

We can think more simply of the mass spectra that arise from the representations of the Lorentz group as possible energy states that a particle can have, just as the hydrogen atom can occupy states of increasing energy. The fact that these states have not yet been experimentally detected could be due to their high energy, far from the energy scale of current experimental techniques. Another interpretation could be that these states with higher spin become strongly interacting at a scale not far above their mass. This means that such particles could arise as composites of other, lower-spin, particles. For instance, in quantum chromodynamics (QCD), there are spin-$\frac{9}{2}$ hadrons that are explained as high-spin excitations of baryons made out of three elementary spin-$\frac{1}{2}$ particles [43, 44].

**6. Concluding discussion**

In the present paper, we have investigated the main algebraic approaches to formulating some of these equations. All the other equations relating to particles with well-defined spin, such as the Dirac, Kemmer–Duffin and Rarita–Schwinger equations, are nothing other than particular cases of one of these generalized equations. The algebraic structure of these equations is the same: what changes is the form of the spin matrices that characterize them. In fact, depending on the initial hypotheses and the physical and mathematical constraints imposed, finite or infinite sets of linear equations are obtained, whose solutions may or may not satisfy the Klein–Gordon equation and which lead to discrete mass spectra that depend on the spin. Each equation of the set is related to a specific representation of the homogeneous Lorentz group, which is unitary only if suitable subsidiary conditions are added to the commutation relations of the spin operators. This suggests that the algebraic apparatus with which the generalized equation should be formulated is perhaps still incomplete, although it is known that the representations of the Lorentz group are deeply connected with the physical nature of particles.

The structure of the mass spectra obtained from the relativistic equations previously described does not reflect the reality of known particles. The symmetries of space-time and the subsidiary conditions imposed by Bhabha, Dirac and Majorana on their theories are not sufficient to construct a model capable of describing particles with any spin. However, by relating the mass energy term that unites these equations with an empirical formula that describes the spectrum of masses of all the particles known to date (in this work, we have chosen that of Sidharth, which is among the most promising in terms of having just a small deviation between theoretical and measured values), it is possible to determine the value of the term $\chi$ that characterizes each of the equations under consideration. This relationship occurs through the mass of the pion, which in the framework we are studying assumes the meaning of physical constant, and that together with the spin value, allows the complete mass spectrum of the known particles to be obtained. It represents the subsidiary condition which permits us to reconsider the *old* relativistic equations within the framework of modern particle physics. With them, in fact, it will be possible to obtain field theories that could disclose the existence of new possible particles that have yet to be observed, or new forms of interaction of matter. Other interpretations of the mass spectra can be given, in which it is assumed that each particle can occupy higher-spin states with energies that are not yet accessible to experimental observation, or in which the high-spin states arise in composite systems with many degrees of freedom.

**Conflict of Interest**

The authors declare that they have no conflicts of interest.